\documentclass[letterpaper]{article} 
\usepackage{aaai25}  
\usepackage{times}  
\usepackage{helvet}  
\usepackage{courier}  
\usepackage[hyphens]{url}  
\usepackage{graphicx} 
\usepackage{natbib}  
\usepackage{caption} 
\usepackage{algorithm}
\usepackage{algorithmic}

\usepackage{newfloat}
\usepackage{listings}
\DeclareCaptionStyle{ruled}{labelfont=normalfont,labelsep=colon,strut=off} 
\floatstyle{ruled}
\newfloat{listing}{tb}{lst}{}
\floatname{listing}{Listing}
\usepackage{booktabs}
\usepackage{multirow}
\usepackage{amsmath}
\usepackage{amsfonts}
\usepackage{pifont}
\newcommand{\tabincell}[2]{\begin{tabular}{@{}#1@{}}#2\end{tabular}}

\title{Phoneme-Level Feature Discrepancies: A Key to Detecting\\ Sophisticated Speech Deepfakes}

\author{
    Kuiyuan Zhang\textsuperscript{\rm 1},
    Zhongyun Hua\textsuperscript{\rm 1}\thanks{Corresponding author},
    Rushi Lan\textsuperscript{\rm 2},
    Yushu Zhang\textsuperscript{\rm 3},
    Yifang Guo\textsuperscript{\rm 4},
}
\affiliations{
    \textsuperscript{\rm 1} Computer Science and Technology, Harbin Institute of Technology Shenzhen, Shenzhen, China\\
    \textsuperscript{\rm 2} Computer Science and Information Security, Guilin University of Electronic Technology, Guilin, China\\
    \textsuperscript{\rm 3} Computer Science and Technology, Nanjing University of Aeronautics and Astronautics, Nanjing, China\\
    \textsuperscript{\rm 4} Alibaba Group, Hangzhou, China\\

    zkyhitsz@gmail.com, huazyum@gmail.com, rslan2016@163.com, yushu@nuaa.edu.cn,guoyifang@gmail.com
}

\usepackage{bibentry}

\begin{document}

\maketitle

\begin{abstract}

Recent advancements in text-to-speech and speech conversion technologies have enabled the creation of highly convincing synthetic speech. While these innovations offer numerous practical benefits, they also cause significant security challenges when maliciously misused. Therefore, there is an urgent need to detect these synthetic speech signals. Phoneme features provide a powerful speech representation for deepfake detection. However, previous phoneme-based detection approaches typically focused on specific phonemes, overlooking temporal inconsistencies across the entire phoneme sequence. In this paper, we develop a new mechanism for detecting speech deepfakes by identifying the inconsistencies of phoneme-level speech features. We design an adaptive phoneme pooling technique that extracts sample-specific phoneme-level features from frame-level speech data. By applying this technique to features extracted by pre-trained audio models on previously unseen deepfake datasets, we demonstrate that deepfake samples often exhibit phoneme-level inconsistencies when compared to genuine speech. To further enhance detection accuracy, we propose a deepfake detector that uses a graph attention network to model the temporal dependencies of phoneme-level features. Additionally, we introduce a random phoneme substitution augmentation technique to increase feature diversity during training. Extensive experiments on four benchmark datasets demonstrate the superior performance of our method over existing state-of-the-art detection methods.


\end{abstract}

%

\section{Introduction}

In today's digital age, advanced machine-learning models have made it increasingly easy to manipulate digital content, raising concerns about the reliability of speech recordings~\cite{mullerDoesAudioDeepfake2022_inthewild}. Speech deepfakes are synthetic recordings that closely mimic a person's speech, making it challenging to verify the authenticity of information~\cite{tanNaturalSpeechEndtoEndTexttoSpeech2024}. Advancements in deep learning technologies for generating realistic speech have made it increasingly difficult to detect these forgeries with conventional methods~\cite{zhangWhatRememberSelfAdaptive2024}.

Deep-learning audio synthesizers typically employ neural networks to replicate the vocal process, often using encoder-decoder architectures to analyze input text and produce synthetic speech~\cite{2024PFlowTTS}. However, authentic human speech production is influenced by complicated acoustic structures and various human bio-parameters~\cite{blue2022whoareyou}. Specifically, acoustic structures, including the lungs, larynx, and articulators, along with human bio-parameters such as gender, health, and age, collaboratively contribute to the complexity of human speech production. These intricate factors present significant challenges for neural synthesizers to replicate these parameters accurately.

\begin{figure}[!t]
    \centering
    \small

    \begin{minipage}[b]{0.9\linewidth}
        \centering
        \begin{minipage}[b]{0.49\linewidth}
            \centerline{\includegraphics[width=1\linewidth]{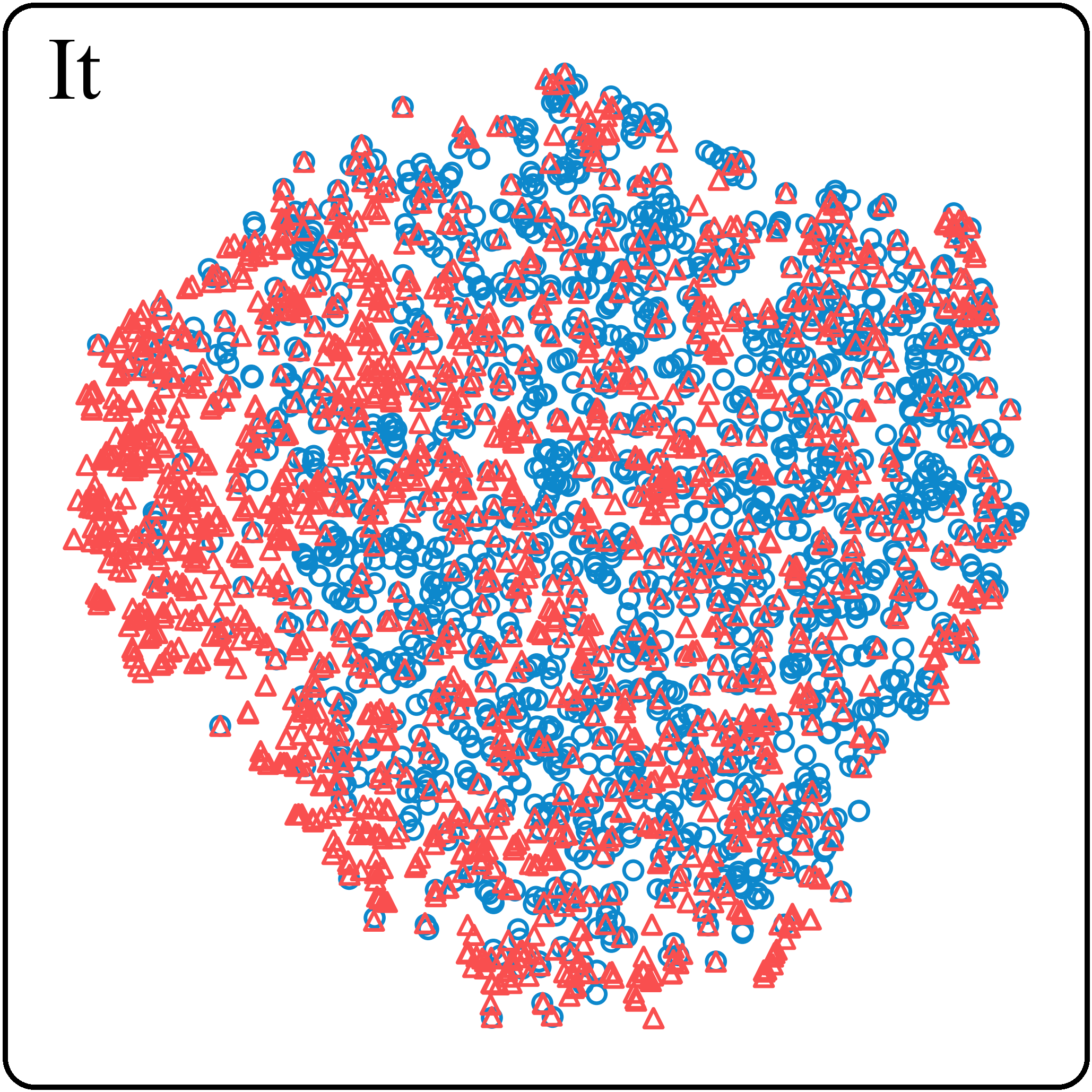}}
        \end{minipage}
        \begin{minipage}[b]{0.49\linewidth}
            \centerline{\includegraphics[width=1\linewidth]{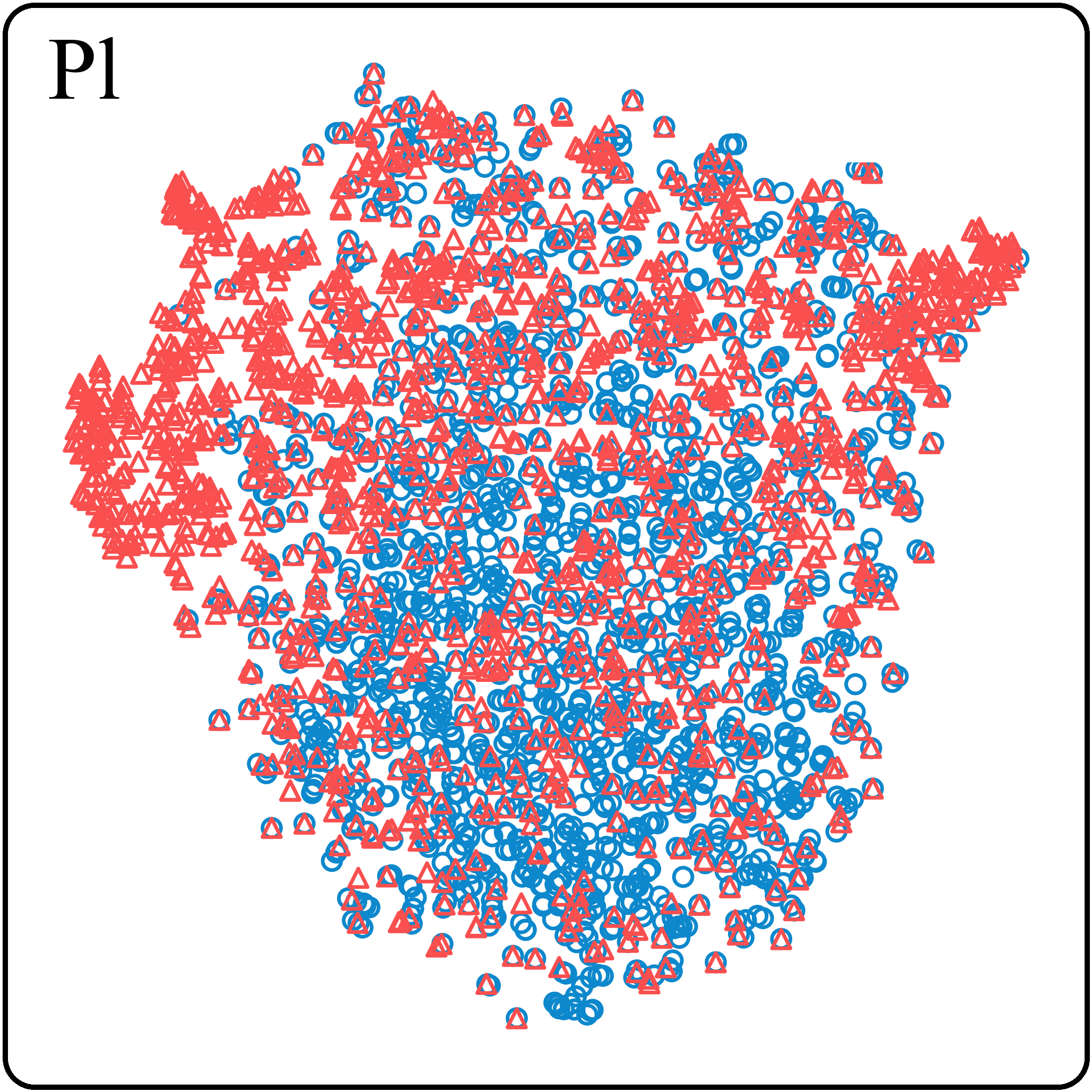}}
        \end{minipage}
        
        \centerline{(a) Frame-level speech features}
        
        \begin{minipage}[b]{0.49\linewidth}
            \centerline{\includegraphics[width=1\linewidth]{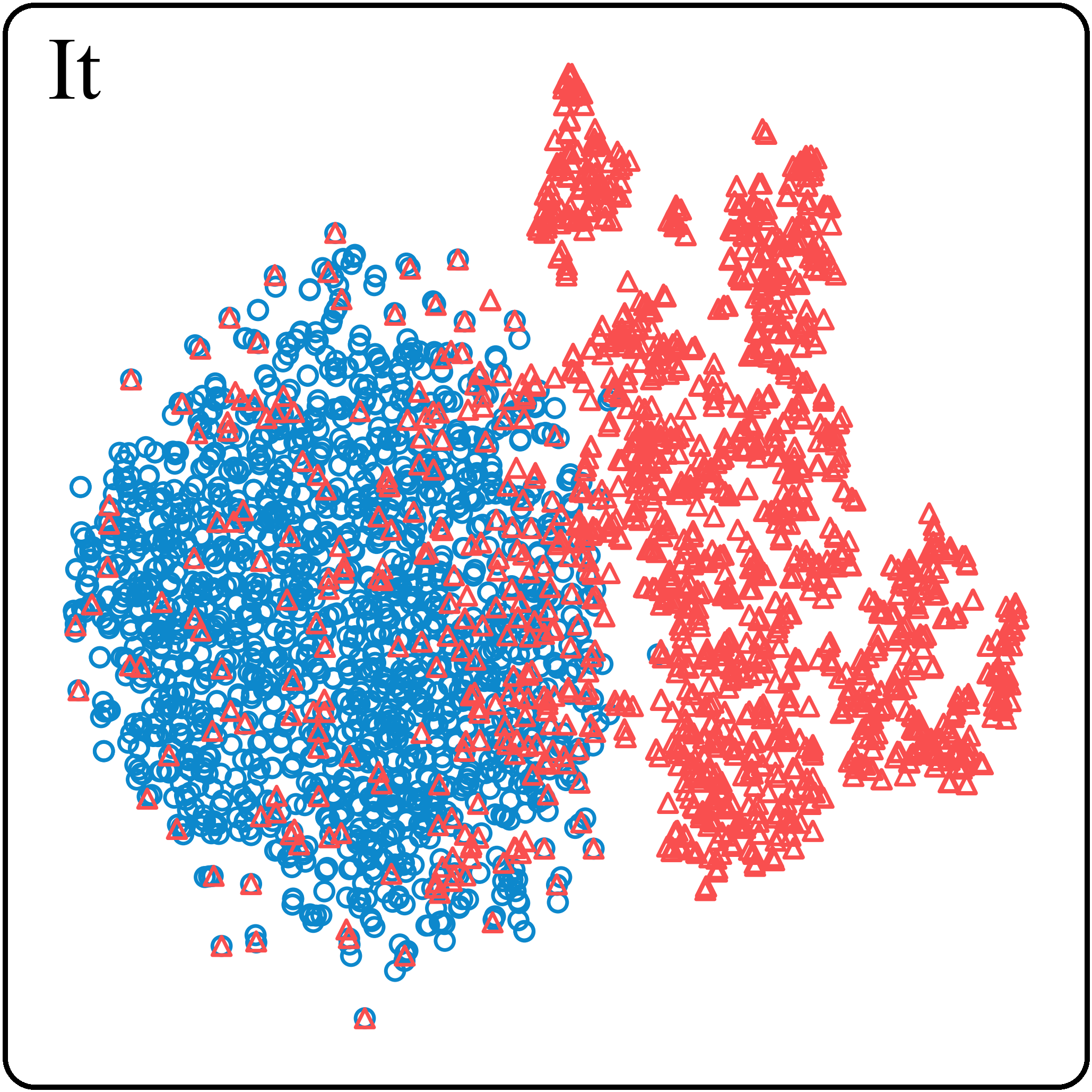}}
        \end{minipage}
        \begin{minipage}[b]{0.49\linewidth}
            \centerline{\includegraphics[width=1\linewidth]{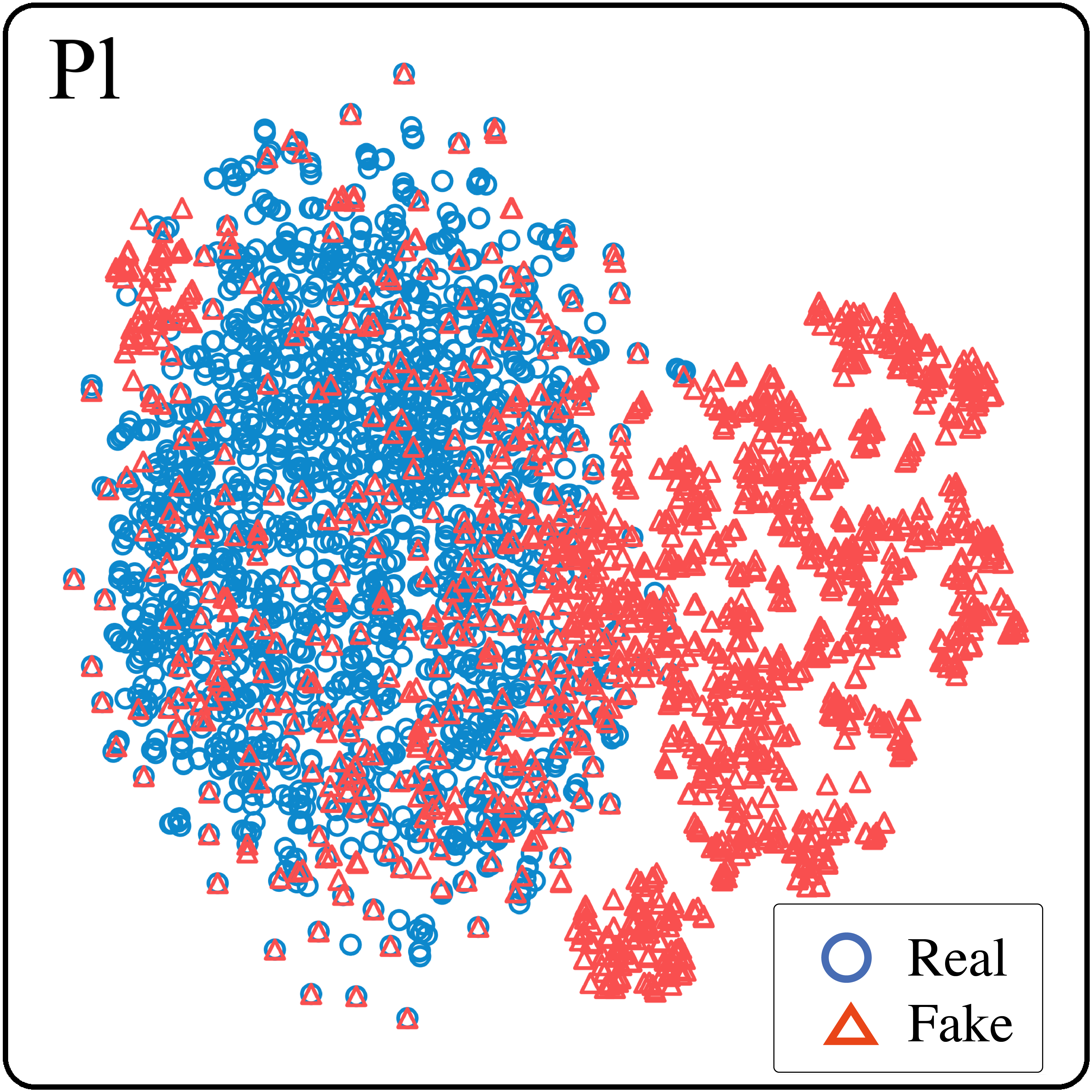}}
        \end{minipage}

        \centerline{(b) Phoneme-level speech features}
        
    \end{minipage}
    
    \caption{T-SNE cluster results. We first employ a pre-trained audio model, Wav2Vec2, to extract the frame-level speech features (last hidden states) from the IT and PL subsets of the MLAAD, a multilingual deepfake speech dataset. The phoneme-level features are generated from frame-level features using adaptive phoneme pooling (see Fig.~\ref{fig:reduce_phoneme}). }
    \label{fig:t_SNE}
    
\end{figure}

\begin{figure}[!t]
    \centering
    \small

    \begin{minipage}[b]{0.99\linewidth}
        \centering
        \centerline{\includegraphics[width=1\linewidth]{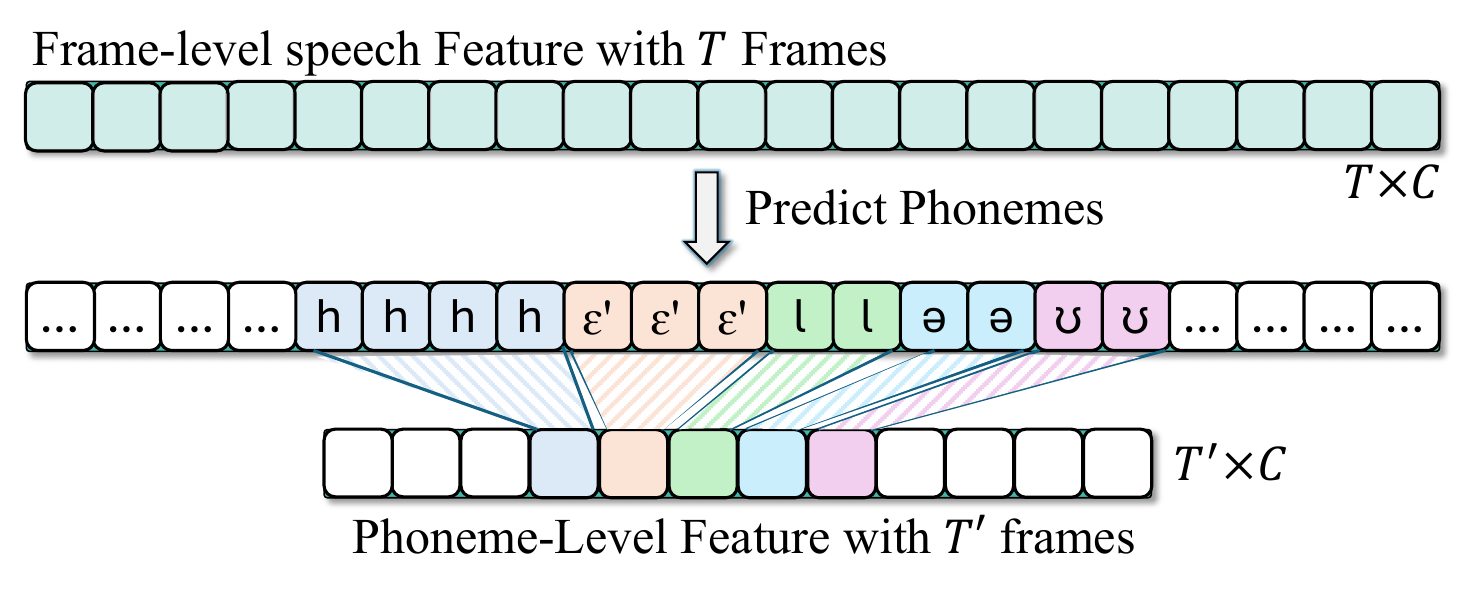}}
    \end{minipage}
    
    \caption{Adaptive phoneme pooling process. Consecutive frames with the same phoneme label in the frame-level feature will combined (averaged) into a vector.}
    \label{fig:reduce_phoneme}
    
\end{figure}

While synthesizers may produce realistic-sounding words, they cannot perfectly simulate the ingredients of words, transitions between words, and the emotional style of the entire sentence. In acoustics, these three elements can be modeled by phonemes, which are the fundamental building blocks of speech. Each phoneme corresponds to a unique configuration of the vocal tract, and transitions between phonemes reflect the individual speaking habits and the sentence styles. Based on this knowledge, previous  studies~\cite{blue2022whoareyou,dhamyal2021using_discover_phonemic_feature} have designed phoneme-based detection models for identifying speech deepfakes. However, these methods often require the extraction of specific phoneme sets for different datasets, making them time-consuming and less generalizable. Additionally, they tend to focus on specific phonemes, ignoring the temporal characteristic of the overall phoneme sequence.

In this work,  we address the limitations of existing phoneme-based methods by focusing on phoneme-level speech features. These features are generated by transforming frame-level speech features into sequences of phoneme features, which preserves both the individual characteristics of each phoneme and the overall sentence style. This approach provides a more accurate representation of vocal tract dynamics and personalized bio-information during speech production. We observe that due to limitations in learning ability and model architecture, current synthesizers are unable to produce perfect and realistic phoneme sequences. This can result in noticeable differences in phoneme-level speech features between deepfake and authentic speech. To validate this observation, we design an adaptive average pooling technique to generate sample-specific phoneme-level features from frame-level features, as shown in Fig.~\ref{fig:reduce_phoneme}.
We then utilize visualization tools to illustrate the effectiveness of these features. As shown in Fig.~\ref{fig:t_SNE}, preliminary validation results confirm that inconsistencies in phoneme-level features between real and fake samples can serve as a reliable indicator for deepfake speech detection.

Building on this, we develop a deepfake audio detection model that relies on these inconsistent phoneme-level speech features. Specifically, we pre-train a phoneme recognition model to predict phonemes, which are then used for adaptive phoneme pooling to produce phoneme-level speech features. We construct edges between adjacent phonemes and employ a graph attention module (GAT) to learn the temporal dependencies in phoneme-level speech features. Additionally, we propose a random phoneme substitution augmentation (RPSA) technique to increase the diversity of speech features during training. Our contributions are:
\begin{itemize}
    \item \textbf{Identifying inconsistent phoneme-level features:} 
    We develop the adaptive phoneme pooling to generate phoneme-level speech features, revealing inconsistencies between authentic and deepfake samples.

    \item \textbf{Constructing a phoneme-based deepfake speech detection model:} 
    We develop a deepfake detection model utilizing a pre-trained phoneme recognition system and GAT, complemented by a data augmentation method.

    \item \textbf{Comprehensive Evaluation:}  
    We conduct extensive experiments across multiple datasets, demonstrating superior performance over state-of-the-art baselines and validating the effectiveness of each component.

\end{itemize}

\begin{figure*}[!th]
    \centering
    \small

    \begin{minipage}[b]{0.5\linewidth}
            \centering
            \centerline{\includegraphics[width=1\linewidth]{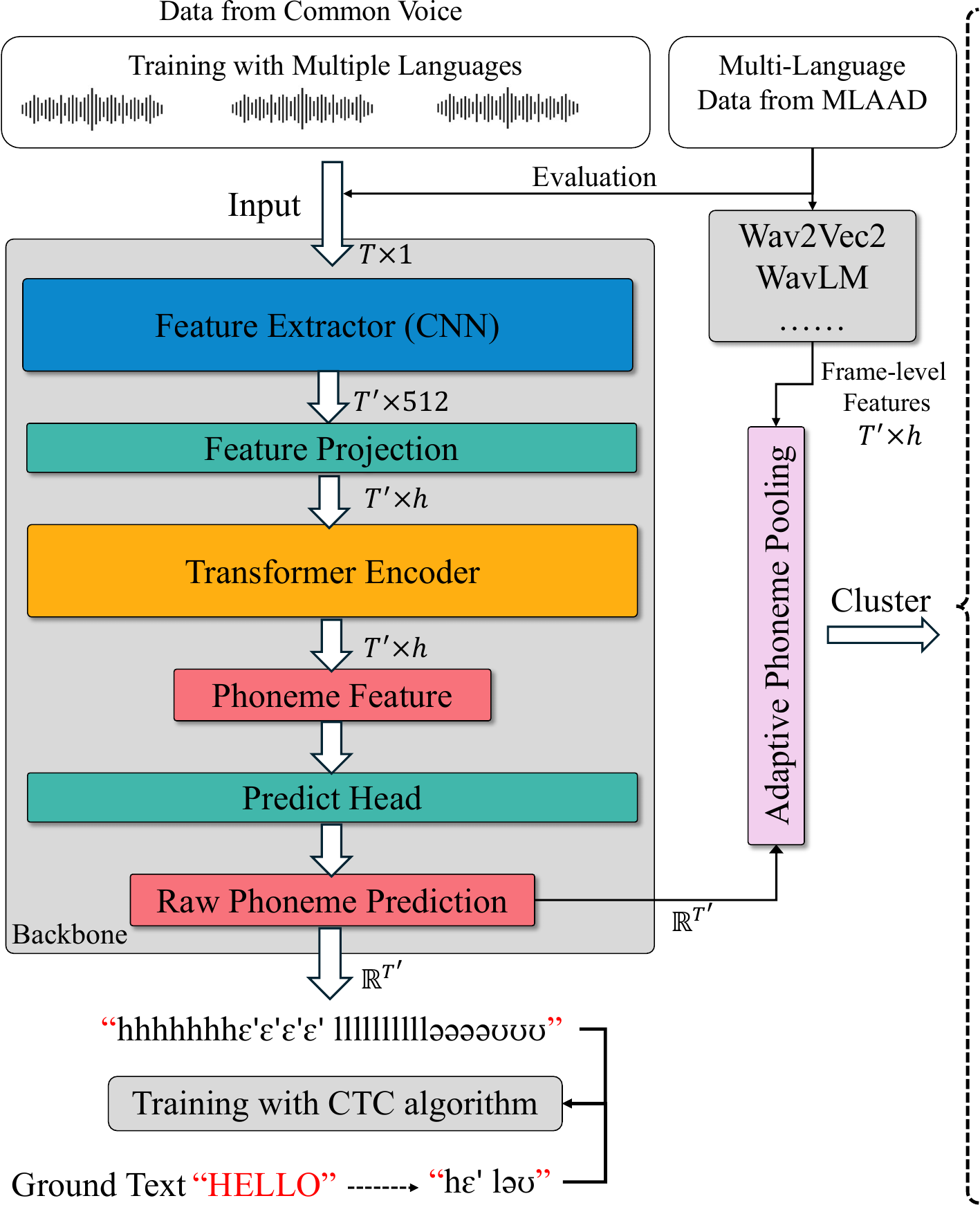}}
    \end{minipage}
    \begin{minipage}[b]{0.4\linewidth}
        \centering
        \begin{minipage}[b]{0.49\linewidth}
            \centerline{\includegraphics[width=1\linewidth]{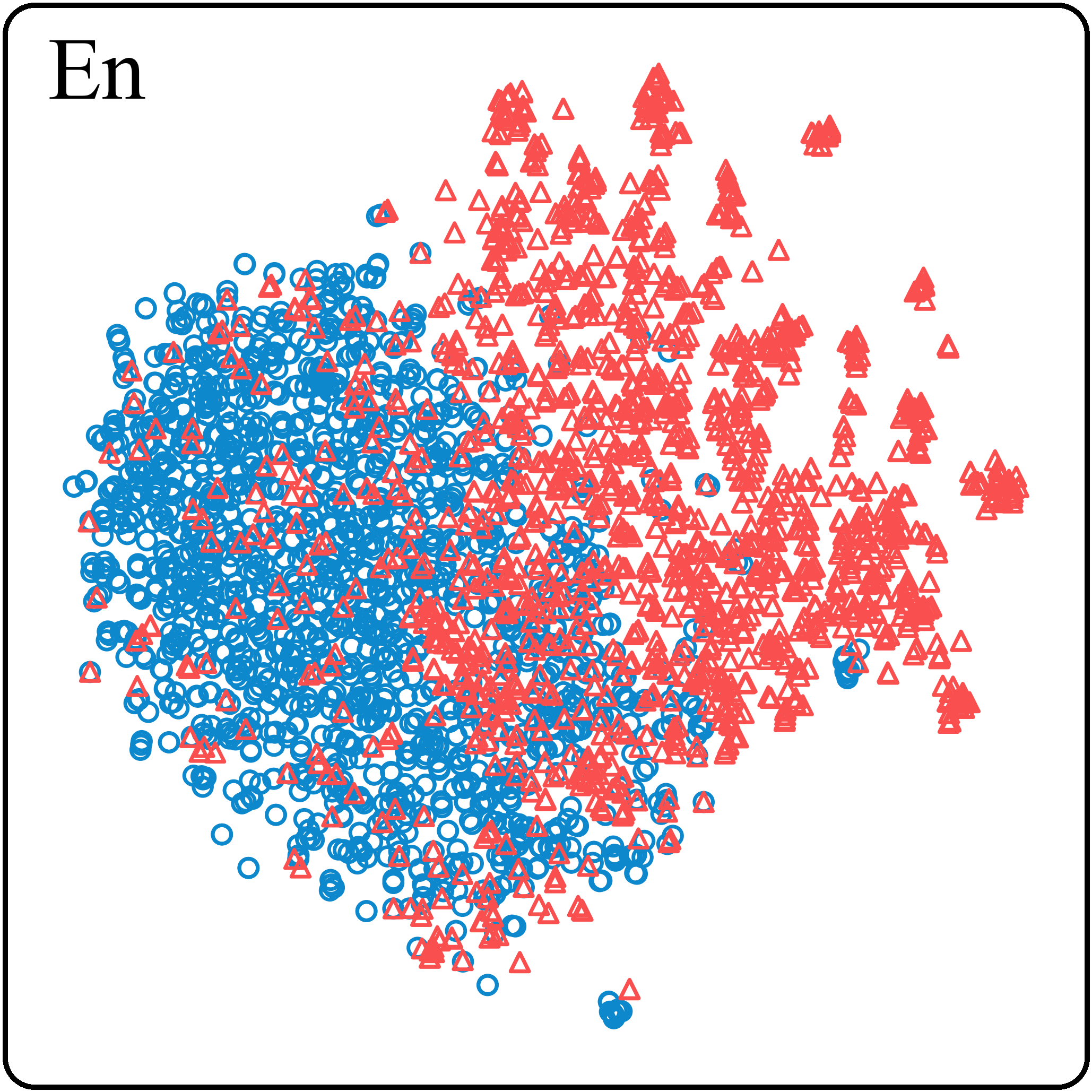}}
        \end{minipage}
        \begin{minipage}[b]{0.49\linewidth}
            \centerline{\includegraphics[width=1\linewidth]{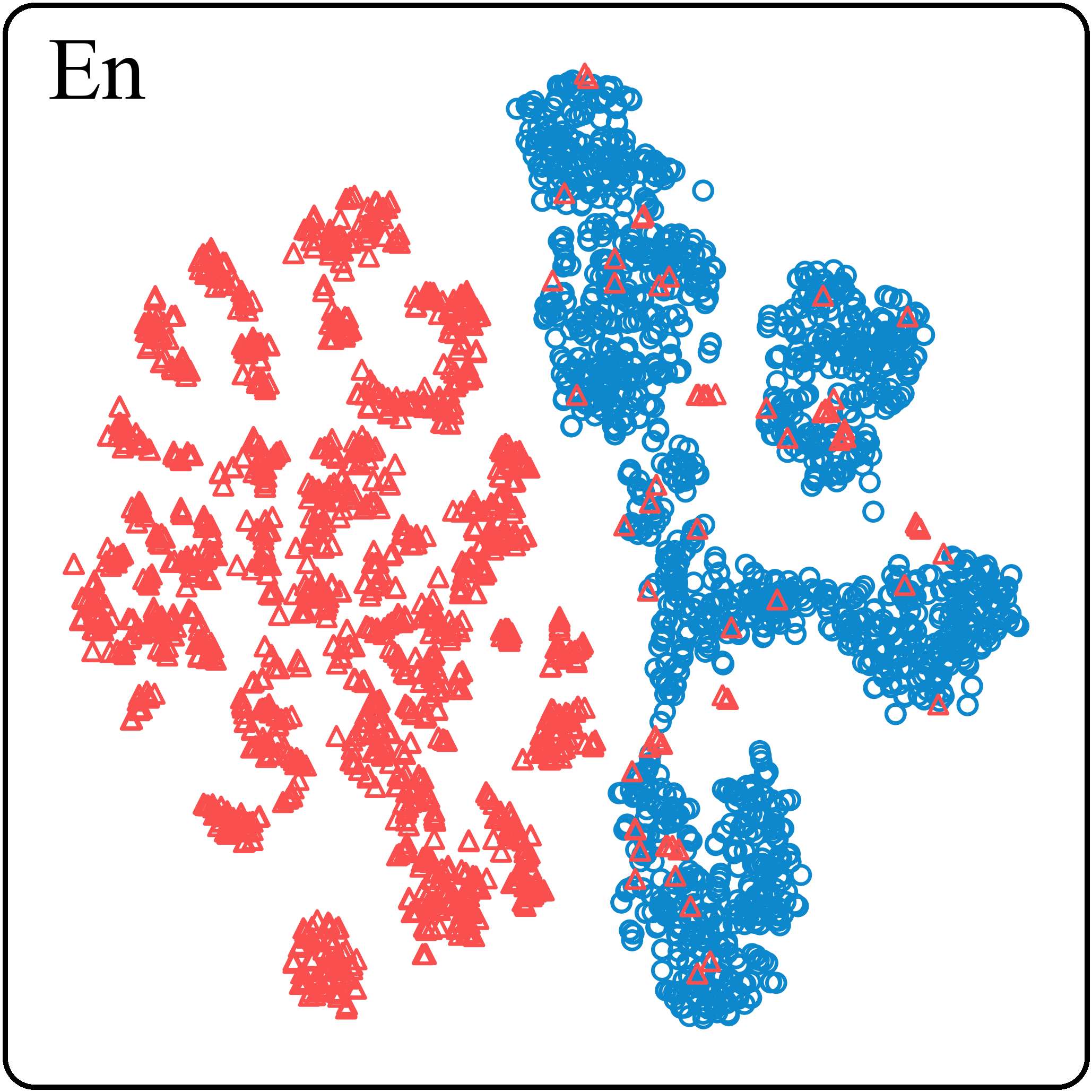}}
        \end{minipage}
        
        \begin{minipage}[b]{0.49\linewidth}
            \centerline{\includegraphics[width=1\linewidth]{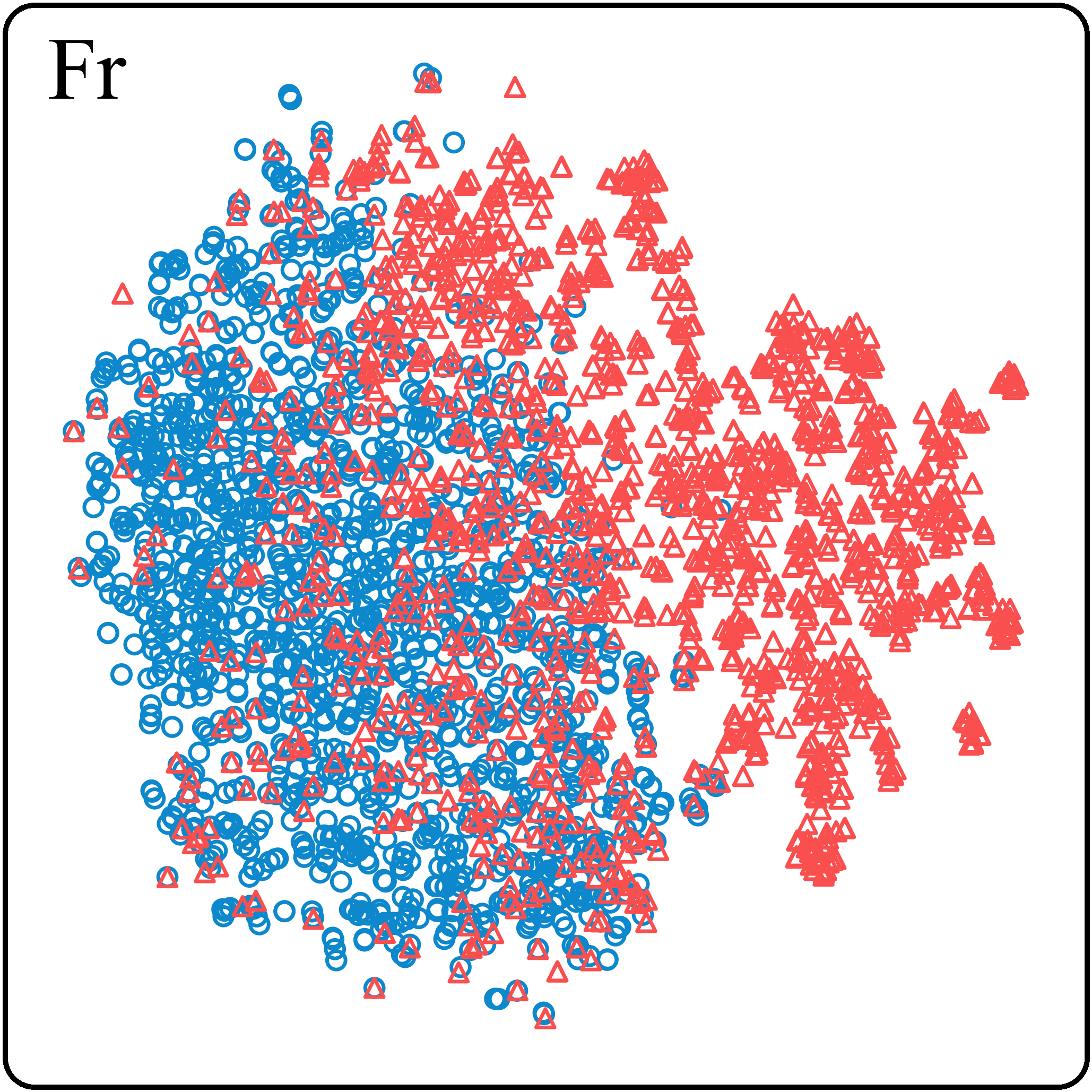}}
        \end{minipage}
        \begin{minipage}[b]{0.49\linewidth}
            \centerline{\includegraphics[width=1\linewidth]{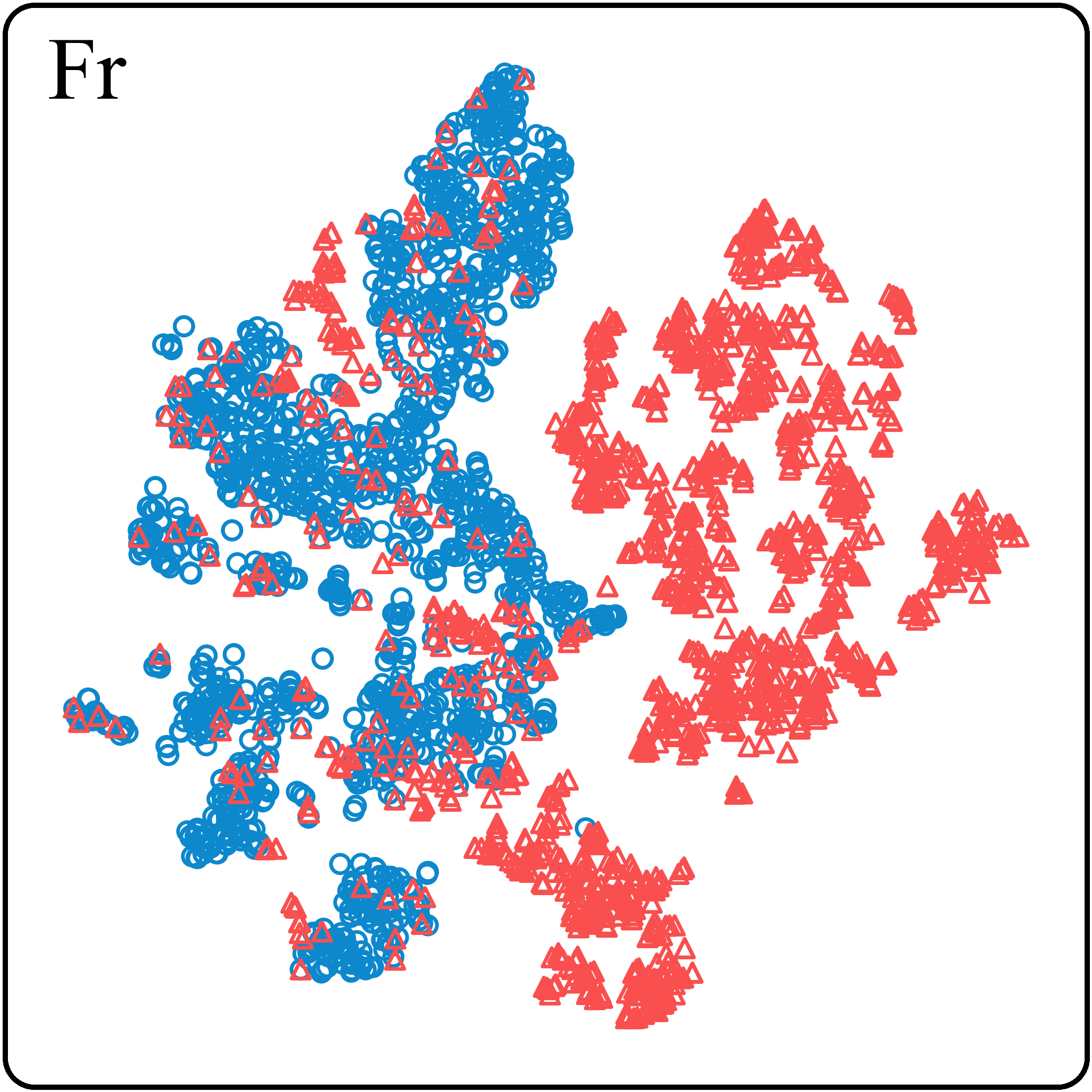}}
        \end{minipage}
        
        \begin{minipage}[b]{0.49\linewidth}
            \centerline{\includegraphics[width=1\linewidth]{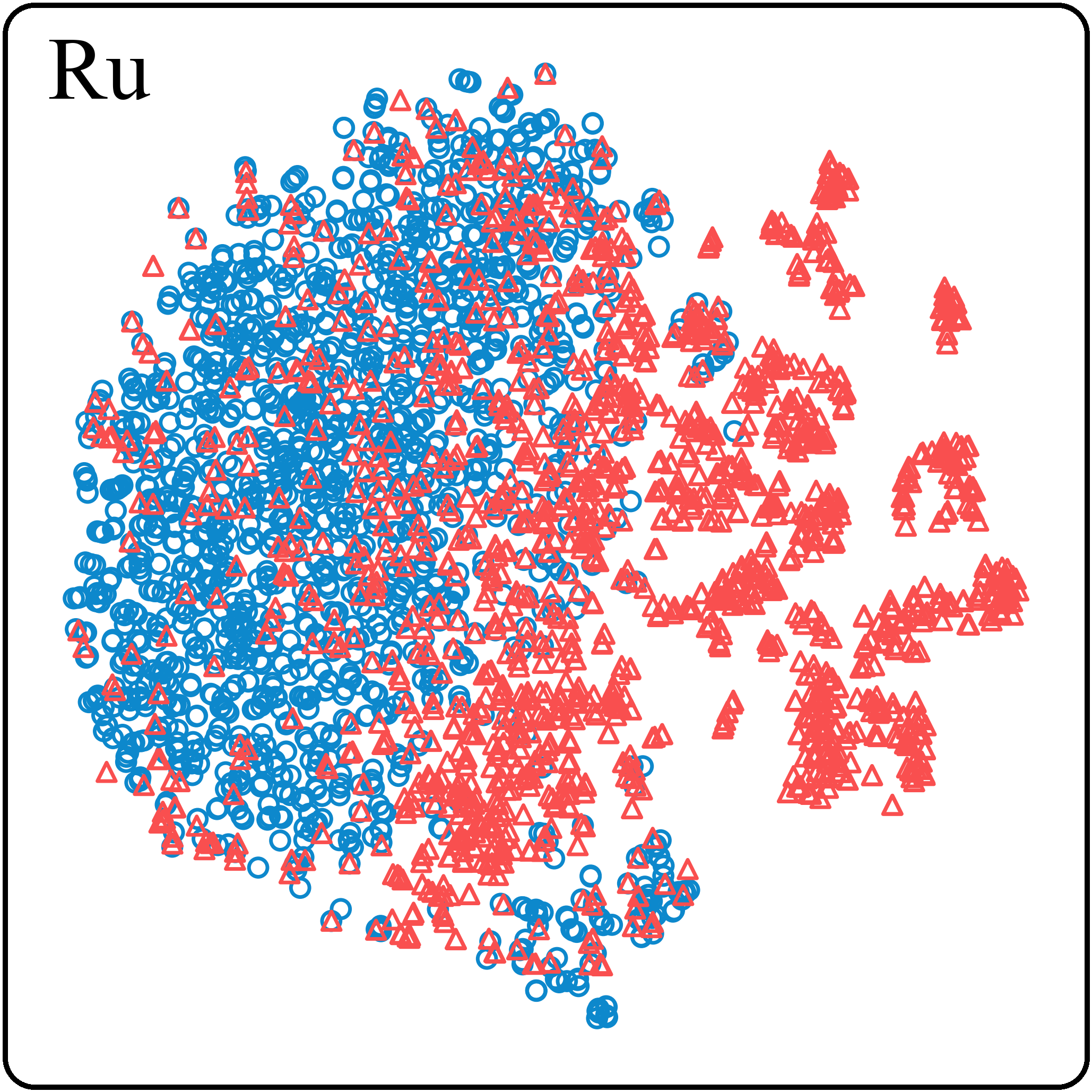}}
        \end{minipage}
        \begin{minipage}[b]{0.49\linewidth}
            \centerline{\includegraphics[width=1\linewidth]{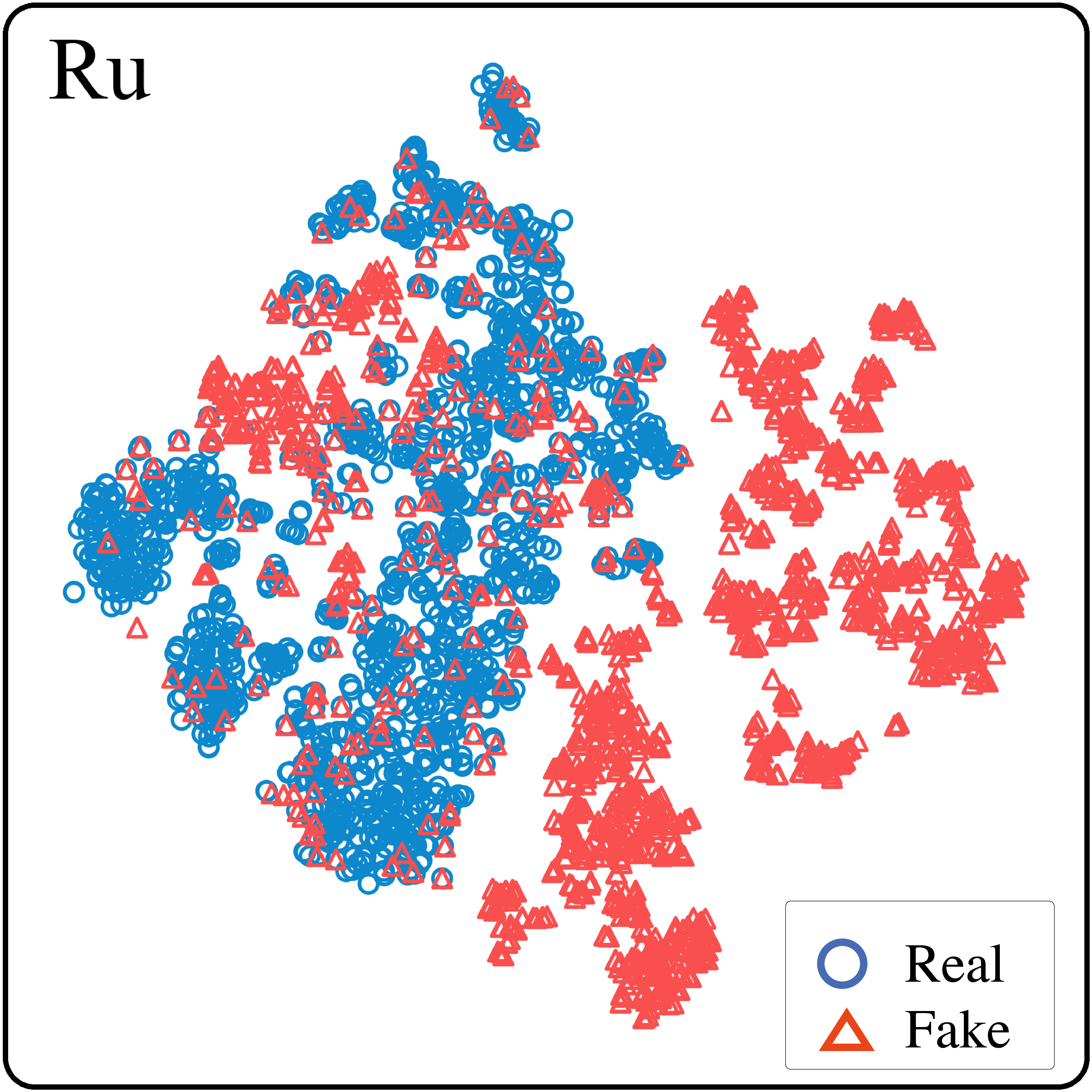}}
        \end{minipage}

        \begin{minipage}[b]{0.49\linewidth}
            \centerline{(a) Wav2Vec2}
        \end{minipage}
        \begin{minipage}[b]{0.49\linewidth}
            \centerline{(b) WavLM}
        \end{minipage}
        
    \end{minipage}
    
    \caption{Training of multilingual phoneme recognition model and T-SNE cluster results of phoneme-level speech features. After training the multilingual phoneme recognition model, we employ it to generate phoneme labels and then generate phoneme-level features from the multilingual frame-level features extracted by Wav2Vec2 and WavLM. T-SNE visualization results demonstrate that phoneme-level features effectively discriminate between real and fake samples.}
    \label{fig:t_SNE_multi_language}
    
\end{figure*}

\section{Related Work}

\subsection{Speech Synthesis}

Speech Synthesis can be produced through Text-to-speech (TTS)~\cite{casanovaYourTTSZeroShotMultiSpeaker2022a} or voice conversion (VC)~\cite{qiPAVITSExploringProsodyAware2024} technologies. TTS methods convert text into speech, while VC methods modify existing speech to change its style. Despite using different inputs, both approaches share a similar \textbf{encoder-decoder} framework.

The encoder processes the input text or speech into embeddings that capture the unique characteristics of the inputs. The decoder then takes these embeddings as input and outputs corresponding speech. In many existing TTS and VC methods~\cite{oord2016wavenet,guan2024mmTTS}, the decoder is further divided into a Mel spectrogram generator and a vocoder. The Mel spectrogram generator produces a Mel spectrogram from the embedding, while the vocoder converts the Mel spectrogram into a synthesized audio waveform. In contrast, those methods~\cite{tanNaturalSpeechEndtoEndTexttoSpeech2024,kimConditionalVariationalAutoencoder2021_VITS} based on conditional variational autoencoder can directly synthesize the waveform from the embeddings.

The rapid development of speech synthesis tools poses increasing challenges to information security.

\subsection{Phoneme-based Deepfake Speech Detection}
The authors in the work~\cite{dhamyal2021using_discover_phonemic_feature}  employed a 2D self-attention model to distinguish spoofed and bonafide speech signals using transformed spectrogram features. By analyzing the attention weights, they discovered that the detection model primarily focuses on only six phonemes in the ASVSpoof2019 Logical Access (LA) Dataset, and the detection performance remains effective even when only the top 16 most-attended phonemes are used as input. Similarly, the work~\cite{blue2022whoareyou} decomposed speech into pairs of phonemes (bigrams) and applied fluid dynamics to estimate the arrangement of the human vocal tract from these bigrams. The whole-sample detection was then conducted by comparing the distribution of bigram features to identify bonafide samples. The authors found that using only 15.3\% phoneme bigrams was sufficient to achieve over 99\% accuracy in their constructed dataset.

While these methods demonstrate the effectiveness of using phonemes for detection, they have certain limitations. They require precise recognition and timestamp labeling of phonemes, which is time-consuming in real applications. Additionally, these approaches utilize individual phonemes or phonemes bigrams extracted from the raw waveform for detection. This makes them ignore the temporal characteristics of the entire phoneme sequence in a more abstract contextual space. As a result, they may miss features that could be useful for deepfake audio detection.



\begin{figure*}[!t]
    \centering
    \small

    \begin{minipage}[b]{0.9\linewidth}
        \centering
        \centerline{\includegraphics[width=1\linewidth]{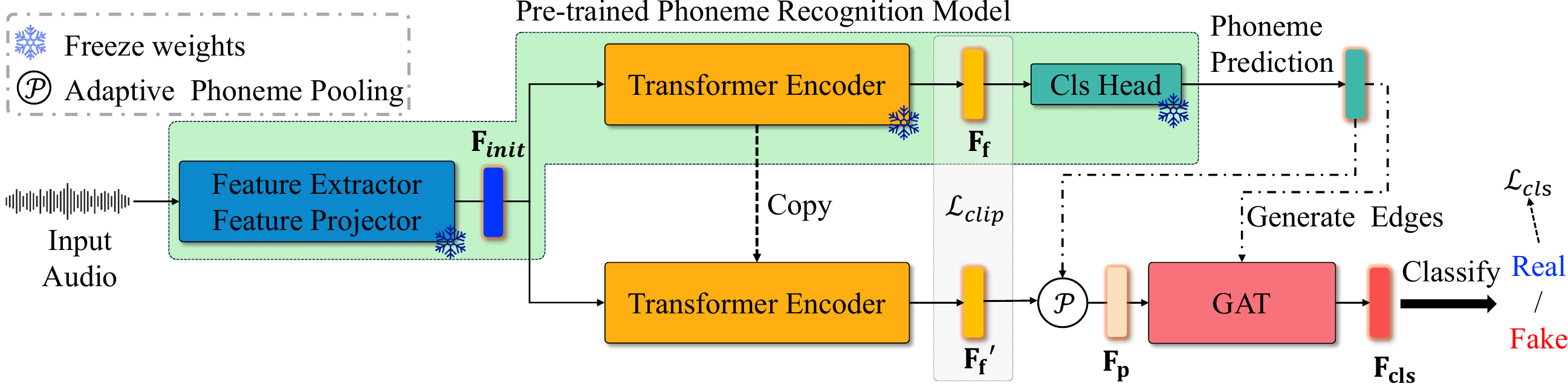}}
    \end{minipage}
    
    \caption{Overview of our deepfake detection model. Given the input feature, our model first uses a pre-trained phoneme-recognition model to predict frame phonemes, then uses a copied Transformer to learn frame-level speech features, next employs GAT to capture temporal dependencies of phoneme-level speech features, and finally makes classification. }
    \label{fig:model}
    
\end{figure*}

\section{Phoneme-level Feature Analysis}

We first pre-train a phoneme recognition model to recognize phonemes for audio frames and then utilize this pre-trained model to generate phoneme-level features for visualization. Note that We only focus on mono audio in this paper. We assume that the input audio is with the shape of $T\times 1$, where $T$ denotes the audio length and the sampling rate is 16k HZ.

\subsection{Pretraining Phoneme Recognition Model}

Generating phoneme-level features requires phoneme labels for each audio frame. In realistic scenarios, phoneme annotations and timestamps are rarely available in audio data, especially for deepfake datasets. Moreover, Deepfake speech is widely available in different language domains. Considering these facts, we train a multilingual phoneme recognition model to recognize phonemes.

\subsubsection{Model Architecture}

In this work, we use a pre-trained audio model, e.g., Wav2Vec2.0~\cite{baevski2020wav2vec} and WavLM~\cite{chen2022wavlm}, as the backbone, which is specifically trained on large-scale audios (English) and can be finetuned on various downstream tasks. As shown in Fig.~\ref{fig:t_SNE_multi_language}, the pretraining audio model consists of a feature extractor, a feature projector and a Transformer encoder. 

The feature extractor and projector employ a 1D CNN to initially extract audio features $\mathbf{F}_{init}$ with a shape of $T' \times h$, where $T'$ is the number of audio frames. The Transformer encoder takes $\mathbf{F}_{init}$ as input and uses self-attention layers to capture the dependencies and correlations in audio frames. We denote the output of the Transformer encoder as the frame-level feature and employ a prediction head for phoneme classification.

\subsubsection{Training}
We adopt the multi-language Common Voice 6.1 corpus to train our phoneme recognition model. Specifically, we select approximately 375k speech samples in 9 languages: English (EN), German (DE), Spanish (ES), French (FR), Polish (PL), Russian (RU), Ukrainian (UK), and Chinese (ZH). Due to computational equipment and time constraints, our final trained multilingual phoneme recognition model has a phoneme error rate (PER) of approximately 0.4 in the validation subset. The full details of the training are presented in the supplementary material.

\subsection{Adaptive Phoneme Pooling}

We design the adaptive phoneme pooling to dynamically generate sample-specific phoneme-level features from frame-level features. The pooling process of it is illustrated in Fig.~\ref{fig:reduce_phoneme}. Assume that the frame-level speech feature has a shape of $T \times C$, where $T$ is the number of audio frames. We first employ a phoneme recognition model to recognize its phoneme labels for each audio frame. Then, we average consecutive audio frames of the same phoneme to generate the phoneme-level speech feature with $T'$ frames ($T' < T$). Since the phoneme lengths and transitions vary according to acoustic structures and personalized bio-parameters, our adaptive phoneme pooling method can generate phoneme-level features with sample-specific characteristics.

\subsection{Visualization of Phoneme-level Features} 

We utilize two public Wav2Vec2 and WavLM models to extract frame-level features from some subsets of the MLAAD dataset and employ the phoneme recognition model to predict phoneme labels for generating phoneme-level features. Note that these two used models were pre-trained on 960 hours of unlabeled English speech samples, thus having no prior access to the MLAAD dataset.

In Fig.~\ref{fig:t_SNE_multi_language}, we present the t-SNE~\cite{vandermaaten2008-t-SNE} cluster results of phoneme-level features. As can be seen from Fig.\ref{fig:t_SNE} and Fig.~\ref{fig:t_SNE_multi_language}, the phoneme-level features reduce the overlap between the real and fake samples in the feature space, resulting in more effective discrimination between the two classes. This separation verifies the inconsistencies of phoneme-level features in fake speech signals, which can be a reliable indicator for deepfake detection. In addition, it can be seen that the phoneme-level features generated by WavLM are more discriminative, which indicates that it can extract more general features than Wav2Vec2 and is a better choice as a backbone.

\section{Deepfake Speech Detector}

\subsection{Overview}

Fig.\ref{fig:model} illustrates the overview model architecture of our detection model, which consists of a frozen pre-trained phoneme recognition model, a copied Transformer encoder, and a GAT module~\cite{velivckovic2017graph-GAT}. Given the input audio sample, the frozen pre-trained phoneme recognition model first uses a feature extractor and projector to extract the initial speech feature $\mathbf{F}_{init}$, following utilizes a Transformer encoder to learn frame-level speech feature $\mathbf{F}_f$, and finally employs a phoneme classification head to predict phonemes. Note that all the parameters in the pre-trained phoneme recognition model are set to be untrainable. 
To detect the deepfake label, we copy and finetune the Transformer encoder to learn frame-level speech feature $\mathbf{F}_f'$ from  $\mathbf{F}_{init}$. Based on the predicted phonemes, we apply average phoneme pooling to $\mathbf{F}_f'$ to obtain the phoneme-level feature $\mathbf{F}_p$. We then employ predicted phonemes to generate edges and utilize the GAT to learn temporal dependency. Finally, we append a classification head for deepfake classification.

\subsection{Graph Attention Module}
We employ the GAT module to capture the temporal dependencies of phoneme-level features  $\mathbf{F}_p\in \mathbb{R}^{T' \times C}$. To do so, we construct edges between consecutive phonemes to model the transition between phonemes. Concretely, assuming $\mathbf{F}_p$ composes $T'$ phoneme vectors $\left\{\vec{f}_1, \vec{f}_2, \ldots, \vec{f}_{T'}\right\}, \vec{f}_i \in \mathbb{R}^C$, we build maximum $N$ edges for each phoneme with its neighborhood phonemes behind: for the $i-$th phoneme where $i \le T'-1$, we add $N-1$ edges $\{i \rightarrow min(i+1, T'), i \rightarrow min(i+2, T'),\cdots,  i \rightarrow min(i+N, T') \}$.

For the $T'$ phoneme vectors in $\mathbf{F}_p$, a graph attention layer (GAL) computes the importance between phoneme vectors using the attention mechanism. The attention coefficients $a_{ij}$ between phoneme $i$ and each of its neighboring phonemes $j\in \mathcal{N}_i$, i.e., the importance of $\vec{f}_j$ to  $\vec{f}_i$, is calculated as:
\begin{equation}
\alpha_{i j}=\frac{\exp \left(\operatorname{LeakyReLU}\left(\overrightarrow{\mathbf{a}}\left[\mathbf{W} \vec{f}_i \| \mathbf{W} \vec{f}_j\right]\right)\right)}{\sum_{k \in \mathcal{N}_i} \exp \left(\operatorname{LeakyReLU}\left(\overrightarrow{\mathbf{a}}\left[\mathbf{W} \vec{f}_i \| \mathbf{W} \vec{f}_k\right]\right)\right)}
\end{equation}
where $\mathbf{W}$ represents the shared linear transformation with weights $\mathbb{R}^{C^{\prime} \times C}$, $\overrightarrow{\mathbf{a}}$ denotes the shared attention transformation with weights $\mathbb{R}^{1\times 2C^{\prime}}$, and the negative input slope in the LeakyReLU nonlinearity is set to $\alpha=0.2$. Then, the final output features for every phoneme $\vec{f}_i$ is calculated as:
\begin{equation}
\vec{f}_i^{\ \prime}=\sigma\left(\sum_{j \in \mathcal{N}_i} \alpha_{i j} \mathbf{W} \vec{f}_j\right)
\end{equation}
where $\sigma$ denotes the exponential linear unit nonlinearity.

In model construction, we stack three GALs and a Long Short-Term Memory (LSTM) layer in the GAT to learn the temporal dependencies sequentially.

\subsection{Deepfake Classification}
Our detection model conducts classification based on the output phoneme-level feature $\mathbf{F}'_p \in \mathbb{R}^{T'\times C}$ of the GAT module. Concretely, we first apply average pooling the temporal dimension of $\mathbf{F}'_p$ and then employ a classification head upon the pooling results  $\mathbf{F}_{cls} \in \mathbb{R}^{C}$ for final classification:
\begin{equation}
    \begin{aligned}
        \hat{y} &= Cls_{head}(Pool_{avg}(\mathbf{F}')).\\
    \end{aligned}
\end{equation}

\subsection{Random Phoneme Substitution Augmentation}

We propose the RPSA technique to improve the feature diversity during training. Specifically, for the extracted feature $\mathbf{F}_{init}^i$ of $i$-th sample in the input batch, we randomly substitute phonemes in $\mathbf{F}_{init}^i$ with the same phonemes of other samples. For instance, if the $k$-th phoneme in $\mathbf{F}_{init}^i$ spans $n_{k}$ frames, it could be replaced by the same phoneme with a possible different number of frames from $\mathbf{F}_{init}^j$. Each phoneme in the sample has a probability of $p$ to be substituted. After obtaining the substituted samples, we collect them into a new batch and feed them into the copied Transformer encoder and GAT to obtain classification results $\hat{y}'$. Note that the label for all the samples in this augmented batch is fake.

\subsection{Loss Function}

We employ the contrastive language-image pre-training (CLIP)~\cite{wu2022wav2clip} loss to increase the semantic similarity between frame-level features $\mathbf{F}_f$ and $\mathbf{F}_f^\prime$. This enables the deepfake detector to also focus on the phoneme prediction task, forming a multi-task learning schedule.
Concretely, the CLIP loss is defined as follows:
\begin{equation}
    \mathcal{L}_{\text{CLIP}} = -\frac{1}{N} \sum_{i=1}^{N} \left( \log \frac{e^{s(g(\mathbf{F}_f^{\prime i}), \mathbf{F}_f^i) / \tau}}{\sum_{k=1}^{N} e^{s(g(\mathbf{F}_f^{\prime i}), \mathbf{F}_f^k) / \tau}} \right)
\end{equation}
where $g$ denotes the a multi-layer perceptron (MLP) to transform $\mathbf{F}_f^\prime$ into another embedding space, $s(a, b)$ represents the cosine similarity function, and $N$ denotes the batch size. Note that the transform $g$ is only used in loss calculation, thus the $\mathbf{F}_f$ and untransformed $\mathbf{F}_f^\prime$ are still in different embedding spaces.

Since audio samples are either bonafide or fake, we use the binary cross-entropy (BCE) loss as the main classification loss: $\mathcal{L}_{cls} = \operatorname{BCE}(\hat{y}, y)$, where $y\in \{0, 1\}$ is the ground truth label of input samples. Besides, the augmentation classification loss is calculated as follows: $\mathcal{L}_{cls}' = \operatorname{BCE}(\hat{y}', \vec{\mathbf{0}})$, where $\vec{\mathbf{0}}$ denotes an all-zero vector. The final optimization objective is defined as follows:
\begin{equation}
    \mathcal{L} = \mathcal{L}_{cls} + 0.5 * (\mathcal{L}_{\text{CLIP}} + \mathcal{L}_{cls}')
\end{equation}

\begin{table*}
    \footnotesize
    \centering
    
    \begin{tabular}{lrrrrrrrrrrrrrrrrr}
    \toprule
    \multirow{2.5}{*}{Method} & \multirow{2.5}{*}{\tabincell{c}{Seen\\ Synthesizers}} & \multicolumn{5}{c}{Unseen Synthesizers} & \multirow{2.5}{*}{\tabincell{c}{Whole\\Testing}} \\ \cmidrule{3-7}
    & & \multicolumn{1}{c}{AR} & \multicolumn{1}{c}{NAR} & \multicolumn{1}{c}{TRD} & \multicolumn{1}{c}{UNK} & \multicolumn{1}{c}{CONC} & \\
    \midrule
LCNN & 91.65 / 16.70 & 85.24 / 24.28 & 83.96 / 25.51 & 92.16 / 16.77 & 89.87 / 19.68 & 90.02 / 18.42 & 87.94 / 21.61 \\
RawNet2 & 87.32 / 21.09 & 81.12 / 27.36 & 82.62 / 26.02 & 87.91 / 18.96 & 82.91 / 24.80 & 85.69 / 22.78 & 83.80 / 24.43 \\
RawGAT & 95.44 / 11.59 & 88.44 / 19.70 & 92.81 / 14.83 & 97.54 / \phantom{0}7.70 & 93.10 / 15.15 & 93.86 / 13.33 & 93.04 / 14.81 \\
LibriSeVoc & 88.66 / 18.23 & 80.45 / 27.74 & 83.87 / 24.81 & 93.70 / 12.94 & 87.28 / 22.84 & 85.69 / 20.00 & 86.25 / 22.86 \\
AudioClip & 90.30 / 18.59 & 83.70 / 24.74 & 81.70 / 26.92 & 90.84 / 17.94 & 84.70 / 24.48 & 88.49 / 20.98 & 85.54 / 23.54 \\
Wav2Clip & 92.53 / 15.48 & 83.26 / 24.82 & 82.88 / 24.75 & 92.47 / 15.42 & 87.36 / 21.37 & 92.86 / 15.21 & 87.72 / 20.98 \\
AASIST & 86.74 / 24.16 & 80.60 / 28.22 & 85.49 / 24.59 & 92.32 / 17.45 & 85.09 / 24.98 & 81.15 / 28.13 & 85.44 / 24.80 \\
MPE & 86.51 / 22.03 & 80.41 / 27.21 & 81.57 / 26.00 & 84.11 / 23.89 & 78.54 / 28.47 & 84.30 / 23.99 & 81.48 / 26.21 \\
ABCNet & 91.68 / 16.47 & 85.58 / 23.12 & 90.52 / 18.35 & 94.53 / 12.69 & 89.77 / 19.65 & 89.85 / 18.20 & 90.08 / 18.87 \\
ASDG & 89.76 / 18.23 & 81.04 / 28.09 & 80.86 / 27.69 & 86.46 / 22.12 & 82.47 / 26.42 & 88.85 / 19.53 & 83.23 / 25.75 \\
Ours & \textbf{99.20} / \phantom{0}\textbf{4.27} & \textbf{97.19} / \phantom{0}\textbf{9.59} & \textbf{98.91} / \phantom{0}\textbf{5.76} & \textbf{99.67} / \phantom{0}\textbf{3.00} & \textbf{97.44} / \phantom{0}\textbf{9.81} & \textbf{99.14} / \phantom{0}\textbf{4.55} & \textbf{98.39} / \phantom{0}\textbf{7.12} \\
    \bottomrule
    \end{tabular}
        \caption{AUC($\uparrow$) / EER($\downarrow$) ($\%$) performances on the ASVspoof2021 DF test subset. All the models are trained and validated on the corresponding training and validation subsets of the ASVspoof2021 DF dataset. }
    \label{tab:ASV2021}
\end{table*}

\begin{table*}[!th]
    \footnotesize
    \centering

    \setlength\tabcolsep{3.5pt}
    
    \begin{tabular}{lcccccccccccccccccc}
        \toprule

        \multirow{2.5}{*}{Model} & \multirow{2.5}{*}{MLAAD Full} & \multicolumn{5}{c}{MLAAD subsets} & \multirow{2.5}{*}{InTheWild} & \\ \cmidrule(lr){3-7}
        & & FR & IT & PL & RU & UK &  \\
        \midrule
LCNN & 96.27 / \phantom{0}9.42 & 98.69 / \phantom{0}6.12 & 97.24 / \phantom{0}9.57 & 99.70 / \phantom{0}2.50 & 89.42 / 19.08 & 98.50 / \phantom{0}5.94 & 30.49 / 64.08 \\
RawNet2 & 85.52 / 22.71 & 86.19 / 21.08 & 83.97 / 24.00 & 94.49 / 13.16 & 94.32 / 12.83 & 82.51 / 26.40 & 71.22 / 33.49 \\
RawGAT & 93.71 / 14.33 & 94.66 / 13.69 & 98.43 / \phantom{0}6.26 & 99.74 / \phantom{0}\textbf{1.53} & 92.91 / 15.42 & 84.76 / 25.08 & 86.85 / 21.72 \\
LibriSeVoc & 83.90 / 23.48 & 80.17 / 28.35 & 87.87 / 20.47 & 96.23 / 11.06 & 92.32 / 15.60 & 88.75 / 17.83 & 66.09 / 36.82 \\
AudioClip & 92.21 / 16.80 & 93.13 / 13.43 & 91.81 / 17.89 & 98.93 / \phantom{0}6.02 & 88.64 / 20.35 & 91.29 / 17.60 & 57.81 / 44.38 \\
Wav2Clip & 94.79 / 12.85 & 97.89 / \phantom{0}5.98 & 99.32 / \phantom{0}\textbf{3.81} & 98.13 / \phantom{0}7.67 & 77.82 / 31.78 & 96.74 / 10.98 & 35.46 / 59.37 \\
AASIST & 92.87 / 14.88 & 93.77 / 14.80 & 98.17 / \phantom{0}5.90 & 99.68 / \phantom{0}1.81 & 90.97 / 18.60 & 83.90 / 27.43 & 83.72 / 25.83 \\
MPE & 94.83 / 12.47 & 93.56 / 14.32 & 96.90 / \phantom{0}9.11 & 97.33 / \phantom{0}8.08 & 87.75 / 20.94 & 96.90 / \phantom{0}8.50 & 69.65 / 35.76 \\
ABCNet & 64.11 / 40.54 & 70.18 / 35.94 & 72.09 / 36.91 & 70.06 / 35.45 & 60.70 / 41.43 & 60.40 / 43.20 & 59.23 / 44.55 \\
ASDG & 94.75 / 10.73 & 97.27 / \phantom{0}8.37 & 97.37 / \phantom{0}7.24 & 99.22 / \phantom{0}3.53 & 81.86 / 28.75 & 96.74 / \phantom{0}7.23 & 26.40 / 69.05 \\
Ours & \textbf{98.88} / \phantom{0}\textbf{5.40} & \textbf{99.43} / \phantom{0}\textbf{3.71} & \textbf{99.33} / \phantom{0}4.00 & \textbf{99.86} / \phantom{0}1.55 & \textbf{99.55} / \phantom{0}\textbf{2.75} & \textbf{98.86} / \phantom{0}\textbf{4.53} & \textbf{91.52} / \textbf{16.07} \\
        \bottomrule
    \end{tabular}
        \caption{AUC($\uparrow$) and EER($\downarrow$) ($\%$) performances on the unseen dataset and languages.}
    \label{tab:MLAAD}
\end{table*}

\section{Experiment Setting}

\subsection{Implementaion Details}

We utilize the WavLM as the backbone of our phoneme recognition model. The number of edges in GAT is set to 10. The substitution probability $p$ in RPSA is set to 0.2. We train our detection model using the AdamW optimizer~\cite{loshchilov2019AdamW}, where the learning rate of the copied Transformer is set to 5e$^{-5}$ and that of other learnable parameters is set to 1e$^{-4}$.

We introduce two data augmentation strategies and an early-stopping technique for every detection approach. Specifically, the data augmentation involves adding random Gaussian noise and applying random pitch adjustments to the audio samples. The early-stopping technique will terminate the training of models if there's no improvement in the area under the Receiver Operating Characteristic (ROC) Curve (AUC) score after three training epochs.
All tests were carried out on a computer equipped with a GTX 4090 GPU, using the PyTorch programming framework.

\subsection{Data}

We utilize the ASVspoof2019~\cite{wang2020asvspoof2019}, ASVspoof2021~\cite{liu2023asvspoof2021}, MLAAD~\cite{müller2024mlaad}, and InTheWild~\cite{mullerDoesAudioDeepfake2022_inthewild} datasets to evaluate our model. The details of these datasets are presented in Table 1 in the supplementary material. For all samples, we randomly clip a three-second clip during the training process and clip the middle three seconds for validation and testing. Audio samples of less than 3 seconds will be padded by itself. Besides, all the samples are converted into 16k HZ.

\subsection{Comparison Methods}

We compare our method with the following detection methods: LCNN~\cite{lavrentyeva2019stc-LCNN}, RawNet2~\cite{jung2020improved-RawNet2}, AASIST~\cite{jung2022aasist}, LibriSeVoc~\cite{sunAISynthesizedVoiceDetection2023-LibraSeVoc}, Wav2Clip~\cite{wu2022wav2clip}, AudioClip~\cite{guzhov2022audioclip}, ABC-CapsNet~\cite{wani2024ABCCapsNet}, MPE~\cite{wang2024multiscale_MPE_LCNN}, ASDG~\cite{xie2024domain_ASDG}. The details of these detection methods are presented in the supplementary material.

\section{Experiment Results}

\subsection{Cross-Method Evaluation}

We conduct the cross-method evaluation on the ASVspoof2021 Deepfake (DF) dataset. Specifically, we train and validate all the methods on the training and validation subsets and split the testing subset into the seen and unseen synthesizer parts. The latter part is further divided into neural vocoder autoregressive (AR), neural vocoder non-AR (NAR), traditional vocoder (TRD), unknown (UNK), and waveform concatenation (CONC).

Table~\ref{tab:ASV2021} shows the evaluation results. As can be seen, our method outperforms the comparison methods significantly in nearly all categories. When tested on seen synthesizers, our method remarkably achieves an AUC of 99.20\% and an EER of just 4.27\%, leading all other models. For unseen synthesizes, our method consistently maintains superior detection performance across all categories, with its best performance of 9.59\%, 5.76\%, 3.00\%, 9.81\% and 4.55\% EER in the AR, NAR, TRD, UNK, and CONC categories. Besides, our method achieves a 7.12\% EER performance for the whole testing subset, significantly outperforming other methods. These experiment results demonstrate that our method has a noticeably enhanced generalization capability in identifying both seen and unseen deepfake methods compared to other models.

\subsection{Cross-Language and Cross-Dataset Evaluation}

In this evaluation task, we train and validate all the detection models on the EN, DE, and ES subsets of the MLAAD dataset and test them on the InTheWild dataset and the rest of the languages of the MLAAD dataset. Table~\ref{tab:MLAAD} shows the cross-language and cross-dataset evaluation results:
\begin{itemize}
    \item \textbf{MLAAD.} When tested on the full testing set, our method can achieve an EER of just 5.40\%, which outperforms other methods. Our method can still obtain superior performance when tested on the FR, RU, and UK subsets.
    \item \textbf{InTheWild.} This dataset consists of collected audio recordings of celebrities and politicians in the real world. One can see that our method achieves the best performance with 16.07\% EER score on the dataset.
\end{itemize}
These cross-evaluation results highlight the superior generalization ability of our method.

\subsection{Robustness Evaluation}

In practical scenarios, audio inputs are rarely pristine. Background noise and varying audio quality commonly exist in real-world audio. We thus test the robustness against random noise and audio compression for each detection model.

\begin{table}[!t]
    \footnotesize
    \centering
    \setlength\tabcolsep{2.5pt}
    \begin{tabular}{lrrrrrrrrrrrrr}
        \toprule
        \multirow{2.5}{*}{Model} & \multirow{2.5}{*}{Seen} & \multicolumn{5}{c}{Unseen Methods} & \multirow{2.5}{*}{Whole} \\ \cmidrule{3-7}
        & & \multicolumn{1}{c}{AR} & \multicolumn{1}{c}{NAR} & \multicolumn{1}{c}{TRD} & \multicolumn{1}{c}{UNK} & \multicolumn{1}{c}{CONC} & \\
        \midrule
LCNN & 25.97 & 30.84 & 32.15 & 25.44 & 27.27 & 28.61 & 29.11 \\
RawNet2 & 20.89 & 27.47 & 26.05 & 19.32 & 24.52 & 22.24 & 24.51 \\
RawGAT & 13.92 & 22.05 & 16.58 & \phantom{0}9.57 & 16.38 & 15.94 & 16.58 \\
LibriSeVoc & 18.79 & 28.15 & 24.91 & 13.21 & 22.53 & 20.67 & 23.18 \\
AudioClip & 19.90 & 26.11 & 27.30 & 19.30 & 25.36 & 21.58 & 24.41 \\
Wav2Clip & 19.76 & 27.63 & 28.48 & 20.70 & 27.15 & 20.34 & 25.13 \\
AASIST & 27.00 & 29.31 & 24.61 & 17.70 & 25.72 & 32.56 & 25.68 \\
MPE & 27.92 & 32.70 & 32.01 & 29.55 & 34.06 & 29.68 & 32.17 \\
ABCNet & 19.41 & 26.05 & 20.67 & 14.92 & 21.65 & 21.75 & 21.21 \\
ASDG & 26.97 & 35.33 & 35.28 & 29.53 & 34.53 & 28.15 & 33.23\\
Ours & \phantom{0}\textbf{5.40} & \textbf{10.80} & \phantom{0}\textbf{6.61} & \phantom{0}\textbf{3.80} & \textbf{11.14} & \phantom{0}\textbf{5.66} & \phantom{0}\textbf{8.17} \\
        \bottomrule
    \end{tabular}
        \caption{Robustness evaluation results (EER$\%$) against background noise on the ASVspoof2021 DF dataset. }
    \label{tab:ASV2021_noise}
\end{table}

\subsubsection{Background Noises}
To assess the robustness against noise, we introduce random background noise during testing. We specifically utilize the Musan~\cite{snyder2015musan} dataset, which provides a broad range of technical and non-technical noises. During testing, we randomly select a noise file from the Musan dataset and add it to the speech sample at a signal-to-noise ratio (SNR) of 20 dB. The robustness evaluation results, presented in Table~\ref{tab:ASV2021_noise}, demonstrate that our method retains a certain level of performance even in the presence of noise.

\begin{table}[!t]
    \footnotesize
    \centering

    \setlength\tabcolsep{1.9pt}
    \begin{tabular}{lrrrrrrrrrrrrr}
        \toprule
        \multirow{2.5}{*}{Model} & \multirow{2.5}{*}{\tabincell{c}{2019\\LA}} & \multicolumn{6}{c}{2021 DF} \\ \cmidrule{3-8}
        & & Whole & \multicolumn{1}{c}{AR} & \multicolumn{1}{c}{NAR} & \multicolumn{1}{c}{TRD} & \multicolumn{1}{c}{UNK} & \multicolumn{1}{c}{CONC} \\
        \midrule
LCNN & 11.37 & 23.90 & 29.56 & 29.67 & 23.25 & 13.30 & 12.56 \\
RawNet2 & 11.38 & 25.31 & 32.36 & 29.10 & 14.61 & 23.88 & 15.50 \\
RawGAT & \phantom{0}4.88 & 20.35 & 26.43 & 22.70 & \phantom{0}8.90 & 17.80 & 17.39 \\
LibriSeVoc & 11.27 & 24.57 & 30.67 & 28.40 & 15.91 & 22.40 & 10.45 \\
AudioClip & 11.05 & 24.19 & 28.16 & 28.67 & 17.99 & 23.02 & 14.52 \\
Wav2Clip & \phantom{0}8.80 & 22.54 & 31.41 & 30.94 & 17.49 & 12.46 & \phantom{0}6.19 \\
AASIST & \phantom{0}4.13 & 19.02 & 26.46 & 20.12 & \phantom{0}9.31 & 17.39 & 14.51 \\
MPE & 15.22 & 30.02 & 33.55 & 30.48 & 26.47 & 32.08 & 21.35 \\
ABCNet & \phantom{0}6.88 & 21.24 & 28.57 & 22.19 & 11.02 & 20.79 & 13.58 \\
ASDG & 12.83 & 25.52 & 31.83 & 32.43 & 25.15 & 14.58 & 16.84 \\
Ours & \phantom{0}\textbf{1.73} & \textbf{10.42} & \textbf{13.74} & \phantom{0}\textbf{9.67} & \phantom{0}\textbf{5.64} & \textbf{12.07} & \phantom{0}\textbf{4.75} \\
        \bottomrule
    \end{tabular}
        \caption{Robustness evaluation results (EER$\%$) against compression artifacts. Models are trained on the ASVspoof2019 LA dataset but tested on the ASVspoof2021 DF dataset. }
    \label{tab:ASV2019_ASV2021}
\end{table}

\subsubsection{Compression Artifacts}
In this task, we train all the models in the ASVspoof2019 LA dataset and then test them in the ASVspoof 2021 DF dataset. Note that the ASVspoof2019 LA dataset does not involve any compression, whereas the samples in the ASVspoof 2021 DF dataset are compressed by various compression methods with different bitrates. The evaluation results, shown in Table~\ref{tab:ASV2019_ASV2021}, indicate that our method achieves the best performance with an EER of 10.42\% when tested on the ASVspoof2021 DF dataset. Furthermore, our method also performs better in the AR, NAR, TRD, UNK, and CONC categories. These experiment results demonstrate that our method exhibits superior robustness against compression artifacts compared to other methods.

\subsection{Ablation Study}

In the ablation studies, we train our method on the MLAAD dataset and report the testing performance on the InTheWild dataset. The ablation results are presented in Table~\ref{tab:ablation}.

\begin{table}[]
    \centering
    \small

    \begin{tabular}{c|ccccccc}
        \toprule
        Setting & GAT& \tabincell{c}{$\mathcal{L}_{CLIP}$} & \tabincell{c}{RPSA} & Pool &  EER \\
        \midrule
        (a) & $\times$ & $\bullet$ &$\bullet$ &$\bullet$ & 28.66 \\
        (b) & $\bullet$ & $\times$&$\bullet$ &$\bullet$ & 26.06 \\
        (c) &$\bullet$ &$\bullet$  & $\times$&$\bullet$ & 22.30 \\
        (d) & $\times$ &$\bullet$ & $\times$ & $\times$ & 32.84 \\
        (f) &$\bullet$ &$\bullet$ &$\bullet$ &$\bullet$ & 16.07 \\
         \bottomrule
    \end{tabular}
        \caption{Ablation studies. The EER (\%) performance on the InTheWild dataset is reported.}
    \label{tab:ablation}
    
\end{table}

\subsubsection{GAT} It learns the temporal dependencies of phoneme-level features. Without it, the EER performance will drop by 12.59\% as shown in setting (a) in Table ~\ref{tab:ablation}, showing the importance of temporal characteristics in deepfake detection.

\subsubsection{$\mathcal{L}_{CLIP}$} It can align the semantic similarity between frame-level features $\mathbf{F}_f$ and $\mathbf{F}_f^\prime$. As shown in setting (a) in Table ~\ref{tab:ablation}, it brings about 10\% EER improvement, proving the necessity of semantic similarity alignment.

\subsubsection{RPSA} The introduction of the RPSA can improve the feature diversity during training. Benefiting from it, our model achieves about 6.2\% improvement in the EER performance, as shown in setting (c) in Table ~\ref{tab:ablation}.

\subsubsection{Adaptive Phoneme Pooling} We develop the adaptive phoneme pooling to extract phoneme-level speech features and use GAT to learn the temporal pattern. Using this pooling method, our model achieves about 16.7\% improvement in the EER performance, as shown in setting (d) in Table ~\ref{tab:ablation}. Note that we also remove the GAT and RPSA in setting (d), since they depend on the adaptive phoneme pooling in training.

\section{Conclusion}
Current deepfake detection methods are increasingly challenged by the rapid advancements in deepfake audio generation. In response, our work introduces a novel approach to deepfake speech detection by focusing on inconsistencies in phoneme-level speech features. We begin by using visualization tools to demonstrate the effectiveness of the phoneme-level feature and subsequently design a deepfake detector based on it. Specifically, by employing adaptive phoneme pooling and a GAT, we effectively capture and analyze phoneme-level features to identify deepfake samples. Additionally, our proposed RPSA technique enhances feature diversity in training. Experimental results demonstrate that our method consistently outperforms state-of-the-art baselines across multiple deepfake speech datasets.

\section*{Acknowledgements}
This work was supported in part by the National Natural Science Foundation of China under Grant 62071142 and by the Guangdong Basic and Applied Basic Research Foundation under Grant 2024A1515012299.

\bibliography{aaai25}

\section{Supplementary Material}

\subsection{Training of Phoneme Recognition Model}

\subsubsection{Loss Function}
The predicted phonemes are longer than the ground truth phonemes since each phoneme can span multiple audio frames. To solve this length mismatch, we adopt the connectionist temporal classification (CTC)~\cite{graves2006connectionist_CTC_loss} loss to train our model. The CTC loss can dynamically map long sequences into short sequences using dynamic programming.

\subsubsection{Evaluation Metric}
We employ the Phoneme Error Rate (PER) as the metric for phoneme recognition pretraining. Concretely, PER denotes the accuracy that quantifies the percentage of errors in the recognized phonemes compared to a reference, which is computed as the following formula:
\begin{equation}
\text{PER} = \frac{Dis_{edit}}{N},
\end{equation}
where $N$ denotes the total number of phonemes in the reference transcription, and $Dis_{edit}$ represents the edit distance, i.e., the number of substitutions, deletions, or insertions to make recognized phonemes same as a reference, and $N$ is the total number of phonemes in the reference transcription. 

\subsubsection{Training Detail}
We employ the Espeak backend to phonemicize the ground truth text of each audio sample. For the collected audio data from the Common Voice\footnote{https://commonvoice.mozilla.org}, we split them at 0.9/0.1 for training and validation. We set the batch size to 16 during training and adopt the AdamW optimizer to optimize the model parameters with the learning rate of 1e-6 and weight decay of 1e-4. We train the model for 50 epochs and save the weights that have the best PER performance on the validation subset.

\subsection{Speech Deepfake Dataset Details}

\begin{table*}[!t]
    \centering
    \caption{Details of the used MLAAD and InTheWild datasets.}
    \label{tab:datasets1}
    \small
    \begin{tabular}{lrrrrrrrrrrrrrrrrrrrrrr}
    \toprule
    \multirow{2}{*}{Details} & \multicolumn{9}{c}{MLAAD}                                   & \multirow{2}{*}{InTheWild} \\ \cmidrule(lr){2-10}
                             & EN      & DE      & ES    & FR & IT & PL & RU & UK & Others &        \\ 
    \midrule
    no. Synthesizers             & 20 & 7 & 5    & 7    & 6    & 5    & 5    & 5    &        8 &   11                 \\
    no. Bonafide             & 31239    & 5856     & 3913 & 5821   &6708    &3808    &3710    &3820    &  0      &   19963               \\
    no. Fake                 & 19000   & 6000    & 4000 & 6000   &7000    &4000    &4000    &4000    &   22000     &    11816                \\
    Total                    & 50239   & 11856   & 7913 &11821    &13708    &7808    &7710    &7820    &  22000      &    23779        \\
    \bottomrule
    \end{tabular}
\end{table*}

\begin{table*}[!t]
    \centering
    \caption{Details of the used ASVspoof2019 LA and ASVspoof2021 DF datasets.}
    \label{tab:datasets2}
    \footnotesize

    \begin{tabular}{lrrrrrrrrrrrr}
    \toprule
    \multirow{2}{*}{Details} & \multicolumn{3}{c}{ASVspoof 2019 LA dataset} & \multicolumn{3}{c}{ASVspoof 2021 DF dataset}  \\  \cmidrule(lr){2-4} \cmidrule(lr){5-7}
                             & Train   & Validation     & Test   & Train   & Validation     & Test  \\
    \midrule
    Synthesizers             & A01-A06 & A01-A06 & A07-A19 & A07-A19 & A07-A19 & \begin{tabular}[r]{@{}r@{}}A07-A19\\ Hub (B00-B01) \\ Hub (D01-D05) \\SPO (N03-N18) \\ Task1 team01-33 \\ Task2 team01-33 \end{tabular}   \\
    no. Bonafide             & 2580    & 2548    & 7355  & 4795    & 973     & 14869 \\
    no. Fake                 & 22800   & 22296   & 63882   & 44530   & 9027    & 65273  \\
    Total                    & 25380   & 24844   & 71237  & 49325   & 10000   & 80142 \\
    \bottomrule
    \end{tabular}

\end{table*}

In our experiments, we utilize the ASVspoof2019~\cite{wang2020asvspoof2019}, ASVspoof2021~\cite{liu2023asvspoof2021}, MLAAD~\cite{müller2024mlaad}, and InTheWild~\cite{mullerDoesAudioDeepfake2022_inthewild} datasets to evaluate detection models. The number of synthesizers and the number of bonafide and fake samples are listed in Table~\ref{tab:datasets1} and Table~\ref{tab:datasets2}. It should be noted that for the  ASVspoof2021 DF dataset, we use full bonafide samples but only partial fake samples in the testing subset. Concretely, we make the number of fake samples match the number of bonafide samples for the five synthesizer categories: neural vocoder autoregressive (AR), neural vocoder non-autoregressive (NAR), traditional vocoder (TRD), unknown (UNK), and waveform concatenation (CONC).

\subsection{Comparison Methods}

We compare our method with the following detection methods: 
\begin{itemize}
    \item LCNN~\cite{lavrentyeva2019stc-LCNN} and RawNet2~\cite{jung2020improved-RawNet2}: They are commonly used baselines in audio detection tasks. LCNN applies a compact 2D CNN to process LFCC features, while RawNet2 directly learns from the waveform using a 1D CNN.
    
    \item AASIST~\cite{jung2022aasist}: It is a graph network that incorporates a heterogeneous stacking graph attention layer to learn speech representations from the raw waveform.

    \item LibriSeVoc~\cite{sunAISynthesizedVoiceDetection2023-LibraSeVoc}: It utilizes RawNet2~\cite{jung2020improved-RawNet2} as the backbone and appends a sub-loss to classify vocoders as an auxiliary task.
    
    \item Wav2Clip~\cite{wu2022wav2clip} and AudioClip~\cite{guzhov2022audioclip}. They are pre-trained models that learn universal speech representations from large-scale speech samples and can adapt to full-stack downstream speech tasks.
    
    \item ABC-CapsNet~\cite{wani2024ABCCapsNet}: This method transforms the input audio into Mel-Spectrogram, utilizes VGGNet~\cite{simonyan2015very_VGGNet} to learn spectrogram features, and finally pioneers the use of cascaded capsule networks to delve deeper into complex speech patterns.
    \item MPE~\cite{wang2024multiscale_MPE_LCNN}: This method designs the multi-scale permutation entropy (MPE) feature and combines it with the LFCC feature as a representation of input audio for classification using an LCNN back-end.
    \item ASDG~\cite{xie2024domain_ASDG}: This method utilizes adversarial learning and triplet loss to learn an ideal feature space that can aggregate real speech and separate fake speech from different domains to achieve better generalizability.
\end{itemize}

\end{document}